\documentclass[superscriptaddress,12pt,preprint, a4paper,nofootinbib]{revtex4}
\usepackage{graphicx}
\usepackage{dcolumn}
\usepackage{bm}
\usepackage{amssymb}
\usepackage{amsmath}
\usepackage{epsfig}    
\usepackage{color}
\def\Fdagger#1{\setbox0=\hbox{$#1$}             
   \dimen0=\wd0                                 
   \setbox1=\hbox{/} \dimen1=\wd1               
   \ifdim\dimen0>\dimen1                        
      \rlap{\hbox to \dimen0{\hfil/\hfil}}      
      #1                                        
   \else                                        
      \rlap{\hbox to \dimen1{\hfil$#1$\hfil}}   
      /                                         
   \fi}

\begin{document}                                                              
\renewcommand{\thefootnote}{\fnsymbol{footnote}}
\preprint{UT-HET-084}
\preprint{KU-PH-013}
\preprint{KEK-TH-1666}

\title{
A UV complete model for radiative seesaw and electroweak 
baryogenesis
based on the SUSY gauge theory
}

\author{Shinya~Kanemura}
\author{Naoki~Machida}
\affiliation{
Department of Physics, University of Toyama, \\ 
3190 Gofuku, Toyama 930-8555, Japan}
\author{Tetsuo~Shindou}
\affiliation{
Division of Liberal Arts, Kogakuin University, \\
1-24-2 Shinjuku, Tokyo 163-8677, Japan}
\author{Toshifumi~Yamada}
\thanks{
Address after August 2013:
Department of Physics and Center for Mathematics and Theoretical Physics, 
National Central University, Chungli, Taiwan 32001, ROC.
}
\affiliation{
KEK Theory Center, \\
1-1 Oho, Tsukuba, Ibaraki 305-0801, Japan}

\begin{abstract}
In a class of supersymmetric gauge theories with asymptotic freedom, 
the low energy effective theory below the confinement scale 
is described by the 
composite superfields of the fundamental representation fields.
Based on the supersymmetric gauge theory with $N_c=2$ and $N_f=3$ with 
an additional unbroken $Z_2$ symmetry, 
we propose a new model where neutrino masses, dark matter, and 
baryon asymmetry of the Universe can be simultaneously explained
by physics below the confinement scale.
This is an example for the ultraviolet complete supersymmetric extension 
of so-called radiative seesaw scenarios with 
first-order phase transition required for successful 
electroweak baryogenesis.
We show that there are benchmark points where all the neutrino 
data, the lepton flavor violation data, and the LHC data
are satisfied. 
We also briefly discuss Higgs phenomenology in this model.
\end{abstract}

\setcounter{footnote}{0}
\renewcommand{\thefootnote}{\arabic{footnote}}
\maketitle

\section{Introduction}

The discovery of the Higgs boson\cite{HiggsLHC} and measurements of 
its properties\cite{HiggsPropLHC} at the LHC provide us a clue to 
explore the essence of electroweak symmetry breaking,
which is possibly described by new physics beyond 
the Standard Model (SM) at the TeV scale.
On the other hand, new physics is required to 
explain phenomena such as neutrino oscillation, 
existence of dark matter (DM) and baryon asymmetry 
of the Universe (BAU). 
If the origins of these phenomena are related to the essence of the Higgs sector, 
they should also arise from new physics at the TeV scale.
In such cases, their origins can be found at current and future 
collider experiments.

For example, let us consider 
the scenario of generating neutrino masses 
by the quantum effect\cite{zee,zeebabu,ma, knt, aks, aks2},
in which tiny neutrino masses are explained by
perturbation of the dynamics at the TeV scale.
There is a class of models with right-handed (RH) neutrinos which are 
assigned the odd-parity under an additional $Z_2$ symmetry\cite{ma, knt, aks, aks2}.
The $Z_2$ symmetry forces the neutrino masses to be generated only at the quantum level,
 giving loop suppression to the neutrino masses.
Also, the lightest $Z_2$-odd particle 
can be a DM candidate if the $Z_2$ symmetry is unbroken.
We call such scenarios as radiative seesaw scenarios.
The Ma model is the simplest one in such a scenario, in which
neutrino masses are generated at the one-loop level by the contribution 
of an extra $Z_2$-odd $\text{SU(2)}_L$ scalar doublet (inert doublet)
and $Z_2$-odd RH neutrinos\cite{ma}.
A neutral component of the inert doublet field or the lightest RH neutrino
can be the DM.
On the other hand, 
in the Aoki-Kanemura-Seto model (the AKS model)\cite{aks,aks2}, where 
$Z_2$-odd charged and neutral singlet scalars as well as $Z_2$-odd 
RH neutrinos are added to a two Higgs doublet
model, neutrino masses are induced at the three-loop level and 
at the same time the lightest $Z_2$-odd particle (the $Z_2$-odd neutral singlet 
or the RH neutrino) can be the DM.
In addition, in this model, 
electroweak baryogenesis can be simultaneously realized 
due to strong first-order electroweak phase transition(1stOPT)
and the CP violating phases in the Higgs sector\cite{EWBG}.

These models of radiative seesaw scenarios have been introduced
as purely phenomenological models.
For example, in the AKS model, some of the coupling constants 
in the Higgs sector and the new Yukawa coupling constants 
for the RH neutrinos are of order one in order to 
satisfy the condition of strong 1stOPT
and also to reproduce the neutrino data.
Consequently, these coupling constants blow up as the energy 
scale increases and the Landau pole appears
at the point much below the Planck scale or the GUT scale,
and the model is well-defined only below the Landau pole\cite{aky}.
This suggests that the model is a low energy effective 
description of a more fundamental theory above the cutoff scale
which corresponds to the Landau pole.
It is then a very interesting question what kind of a fundamental theory can
lead to such a low energy effective theory.

In this paper, we propose a concrete model of the
fundamental theory whose low-energy description gives a phenomenological model\cite{aks} of 
radiative seesaw scenarios with electroweak baryogenesis.
In this model, the origin of the Higgs force above the cutoff scale of the low energy theory
is a new gauge interaction with asymptotic freedom.
In order to describe this picture, we consider 
the supersymmetric (SUSY) SU($N_c$) theory with $N_f$ flavors\cite{fatHiggs,fatbutlight}.
For $N_f=N_c+1$,
confinement occurs at an infrared (IR) scale $\Lambda_H$\cite{IntrSei}.
We here consider the simplest case with $N_c=2$ and $N_f=3$\footnote{
	This is the same choice as in the minimal SUSY fat Higgs model\cite{fatHiggs}.
	In this model, however, 
	additional heavy superfields are introduced
	in order to make 
	some of the unnecessary composite superfields to be very heavy.
	Consequently,
	in the low energy effective theory of the model, 
	two $\text{SU}(2)_L$ doublet and one singlet Higgs superfields 
	appear as composite states of fundamental
	superfields of the $\text{SU}(2)_H$ gauge symmetry, corresponding to
        the field content of the nearly-minimal SUSY SM (nMSSM)\cite{nMSSM}.
}.
In the low-energy effective theory below the confinement scale $\Lambda_H$,
Higgs superfields $H_{ij}(\sim T_iT_j)$ appear as the composite states 
of the fundamental superfields $T_i\;(i=1,\cdots,6)$ which are doublets of 
the $\text{SU}(2)_H$ gauge symmetry\cite{composite}.
In order to realize radiative seesaw scenarios in the low-energy effective theory,
we add elementary RH neutrino superfields $N_i^c\;(i=1,\cdots 3$) to the 
model.
We further impose a $Z_2$ symmetry to the model assuming that $N_i^c$ and some of 
$T_i$'s are $Z_2$-odd.
Below the confinement scale, the symmetries of the model are
$\text{SU}(3)_C\times \text{SU}(2)_L\times \text{U}(1)_Y\times Z_2$, under which 
fifteen Higgs superfields appear\cite{KSYam}.
All the scalar particles required in the AKS model are included in
these fifteen Higgs superfields. 
It is quite interesting that the complicated particle content of
the AKS model is predicted by this 
$\text{SU}(2)_H\times Z_2$ theory above the cutoff scale 
without any artificial assumption.

The condition of strongly 1stOPT,
$\varphi_c/T_c\gtrsim 1$,
which is required for successful electroweak baryogenesis 
determines the size of the coupling constants of the Higgs potential at the electroweak scale.
This property commonly results in the enhanced triple Higgs boson constant~\cite{KOS,GroSer}.
The electroweak baryogenesis scenario can partially be tested by measuring the triple Higgs boson coupling 
at future collider experiments. 
By the renormalization group equation (RGE) analysis of the coupling constant, the scale of 
the Landau pole is evaluated as $\mathcal{O}(10)\text{TeV}$\cite{KSS,KSSY},
which is identical to the confinement scale $\Lambda_H$ under the 
Na$\ddot{\i}$ve Dimensional Analysis (NDA)\cite{nda}.

In our model, 
the lightest $Z_2$-odd particle in the 
effective theory can be a DM candidate as in usual radiative 
seesaw scenarios.
If the $R$-parity is also imposed, 
there are two discrete symmetries, and 
a rich possibility for the multi-component DM scenario
occurs\cite{multiDM}. 
In this paper, however, we do not specify the scenario of 
DM. Detailed analysis for the multi-component DM scenario will be performed 
in our model elsewhere\cite{future}.

We show that the neutrino masses are generated at the loop level
in the low-energy effective theory of our model.
It contains diagrams of both the Ma model and the AKS model.
We find benchmark points in the parameter space where all the current
experimental data for Higgs bosons, neutrino data, 
constraints of lepton flavor violation processes
and the condition of strongly 1stOPT
are satisfied.
We also discuss the possibility of testing this model at current 
and future collider experiments.

\section{Model based on SUSY strong dynamics}

In this section, we will briefly review a SUSY model with $\text{SU(2)}_H\times Z_2$ 
symmetry and six chiral superfields, denoted by $T_i \ (i=1,\cdots, 6)$, 
which are doublets of the $\text{SU}(2)_H$ gauge symmetry.
The superfields $T_i$'s are also charged under the SM gauge groups $\text{SU}(2)_L \times U(1)_Y$.
The SM charges and $Z_2$ parity assignments on $T_i$'s are given in Table~\ref{Tfields}.
\begin{table}
\caption{
The SM charges and $Z_2$ parity assignments on the $\text{SU}(2)_H$ doublets $T_i$.
\label{Tfields}
}
\begin{center}
\begin{tabular}{|c|c|c|c|c|c|} \hline
	Superfield                 & $\text{SU}(3)_C$& $\text{SU}(2)_L$ & $\text{U}(1)_Y$&$Z_2$ \\ \hline
$\left(
\begin{array}{c}
T_1  \\
T_2
\end{array}
\right)$          &1    & 2           & 0             &$+1$   \\ \hline
$T_3$             &1  & 1           & +1/2            &$+1$ \\ \hline
$T_4$             &1  & 1           & $-$1/2          &$+1$ \\ \hline
$T_5$             &1  & 1           & +1/2            &$-1$ \\ \hline
$T_6$             &1  & 1           & $-$1/2          &$-1$ \\ \hline
\end{tabular}
\end{center}
\end{table}
The tree-level superpotential respecting all the gauge symmetries and the $Z_2$ parity
 is written as
\begin{align}
	W_{\text{tree}}  &=  m_{1} T_{1} T_{2}  +  m_{3} T_{3} T_{4}  +  m_{5} T_{5} T_{6}\;.
\end{align}

The SU(2)$_H$ gauge coupling becomes non-perturbative at an IR scale, denoted by $\Lambda_H$. 
Below the scale $\Lambda_H$,
the theory is described in terms of 
composite chiral superfields, $H_{ij}^{\prime} = T_{i}T_{j} \ (i \neq j)$, which are singlets of SU(2)$_H$.
We have the following dynamically generated superpotential below $\Lambda_{H}$:
\begin{align}
	W_{\text{dyn}} &= -\frac{1}{\Lambda^3} \epsilon^{ijklmn}  H_{ij}^{\prime}  H_{kl}^{\prime}  H_{mn}^{\prime}\;,
\end{align}
where $\Lambda$ is a dynamically generated scale\cite{IntrSei}.
The total effective superpotential is 
simply the sum of $W_{\text{dyn}}$ and $W_{\text{tree}}$:
\begin{align}
	W_{\text{eff}} &= W_{\text{dyn}}  +  W_{\text{tree}}
= W_{\text{dyn}}  +  m_{1} H_{12}^{\prime}   +  m_{3} H_{34}^{\prime}   +  m_{5} H_{56}^{\prime} \;.
\end{align}
We cannot determine the normalization for the dynamically generated superpotential.
The effective K\"{a}hler potential below the scale $\Lambda_H$
is also undetermined, and so is the canonical normalization 
for the mesonic superfields.
However,the NDA suggests the following form of the effective K\"{a}hler potential
and normalization for the effective superpotential at the scale $\Lambda_H$\cite{nda}:
\begin{align}
	K_{\text{eff}}[\Lambda_H] &\simeq  \frac{1}{16 \pi^2 \Lambda_H^2}  H_{ij}^{\prime\dagger} H_{ij}^{\prime}\;,
\\
W_{\text{eff}}[\Lambda_H] &\simeq   -\frac{1}{16 \pi^2 \Lambda_H^3} 
\epsilon^{ijklmn} H_{ij}^{\prime}  H_{kl}^{\prime}  H_{mn}^{\prime}
 +  m_{1} H_{12}^{\prime}  + m_{3} H_{34}^{\prime}  +  m_{5} H_{56}^{\prime}\;.
\end{align}
The canonically normalized mesonic superfields $H_{ij}$ at the scale $\Lambda_H$
 are then given by
\begin{align}
H_{ij}  &\simeq  \frac{1}{4\pi \Lambda_H} H_{ij}^{\prime}\;,
\end{align}
 and the superpotential at the scale $\Lambda_H$ is rewritten as
\begin{align}
	W_{\text{eff}}[\Lambda_H] \simeq  4\pi \epsilon^{ijklmn} H_{ij}  H_{kl}  H_{mn}
							  +  4\pi \Lambda_H m_{1} H_{12}  +  4\pi \Lambda_H m_{3} H_{34} 
 +  4\pi \Lambda_H m_{5} H_{56}\; .
\label{eff superpotential}
\end{align}

The basic setup explained above is the same as the one in the minimal SUSY fat Higgs model\cite{fatHiggs}.
In general, fifteen mesonic superfields $H_{ij}$ appear in the low-energy effective theory of 
the fundamental $\text{SU}(2)_H$ gauge theory with 3 flavors.
In the minimal SUSY fat Higgs model, the superfields in the low-energy effective theory 
are made to be identical to those in the nMSSM\cite{nMSSM} by introducing several $\text{SU}(2)_H$ singlet 
superfields which give masses as large as $\Lambda_H$ to ten of the 
fifteen mesonic superfields.
On the other hand, in our model, we do not introduce such additional singlets and
thus all the fifteen mesonic chiral superfields remain in the effective theory below $\Lambda_H$.

We identify the fifteen mesonic chiral superfields, $H_{ij}$,
with the MSSM Higgs doublets, $H_u, \ H_d$, and the exotic chiral superfields
in an extended Higgs sector, as
\begin{align}
H_{u} &\equiv
\left(
\begin{array}{c}
H_{13}  \\
H_{23}
\end{array}
\right)\;,\quad
H_{d} \equiv
\left(
\begin{array}{c}
H_{14}  \\
H_{24}
\end{array}
\right)\;,\quad
\Phi_{u} \equiv
\left(
\begin{array}{c}
H_{15}  \\
H_{25}
\end{array}
\right)\;,\quad
\Phi_{d} \equiv
\left(
\begin{array}{c}
H_{16}  \\
H_{26}
\end{array}
\right)\;,\quad \nonumber
\\
N &\equiv H_{56}\;,\quad
N_{\Phi} \equiv H_{34}\;,\quad
N_{\Omega} \equiv H_{12}\;,\quad \nonumber
\\
\Omega^{+} &\equiv H_{35}\;,\quad
\Omega^{-} \equiv H_{46}\;,\quad
\zeta \equiv H_{36}\;,\quad
\eta \equiv H_{45}\;.\quad
\end{align}
The SM charge and $Z_2$ parity of these Higgs superfields are summarized in Table~\ref{extraHiggsTable}.
\begin{table}
\caption{
The field contents of the Higgs sector below $\Lambda_H$.
\label{extraHiggsTable}
}
\begin{center}
\begin{tabular}{|c|c|c|c|c|} \hline
	Field        &$\text{SU}(3)_C$ & $\text{SU}(2)_{L}$ & $\text{U}(1)_{Y}$ &$Z_2$\\ \hline
	$H_{u}$      &1         & 2           & +1/2 		&$+1$           \\ \hline
	$H_{d}$      &1         & 2           & $-$1/2   	&$+1$         \\ \hline
	$\Phi_{u}$   &1         & 2           & +1/2		&$-1$             \\ \hline
	$\Phi_{d}$   &1         & 2           & $-$1/2	&$-1$              \\ \hline
	$\Omega^+$   &1            & 1           & +1   	&$-1$         \\ \hline
	$\Omega^-$   &1            & 1           & $-$1   &$-1$       \\ \hline
	$N$, $N_{\Phi}$, $N_{\Omega}$&1        & 1           & 0      &$+1$        \\ \hline
	$\zeta$, $\eta$   &1     & 1           & 0      &$-1$        \\ \hline
\end{tabular}
\end{center}
\end{table}
With these fields, the superpotential in Eq.~(\ref{eff superpotential}) is rewritten as
\begin{align}
	W_{\text{eff}} &= \hat{\lambda} \left\{ N (H_{u}H_{d} + v_{0}^{2})  +  N_{\Phi} (\Phi_{u}\Phi_{d} + v_{\Phi}^{2}) 
 +  N_{\Omega} (\Omega^{+}\Omega^{-} + v_{\Omega}^{2}) \right. \nonumber
\\
&- \left.  N N_{\Phi} N_{\Omega}  -  N_{\Omega} \zeta \eta  + 
\zeta H_{d} \Phi_{u}  +  \eta H_{u} \Phi_{d}
 -  \Omega^{+} H_{d} \Phi_{d}  -  \Omega^{-} H_{u} \Phi_{u}  \right\} \; .
\end{align}
After the $Z_2$-even neutral fields $N$, $N_{\Phi}$, and $N_{\Omega}$ get vacuum expectation values (vev's), 
the relevant terms of the effective superpotential are given by\cite{KSYam, KSSY}
\begin{align}
	W_{\text{eff}} =& -\mu H_uH_d-\mu_{\Phi}\Phi_u\Phi_d-\mu_{\Omega}(\Omega^+\Omega^- - \zeta\eta)
	\nonumber\\
	&+\hat{\lambda} \left\{
	H_d\Phi_u\zeta + H_u\Phi_d\eta - H_u\Phi_u\Omega^- - H_d\Phi_d\Omega^+
\right\}\;.
\label{eq:superpotentialWeff}
\end{align}
The relevant soft SUSY breaking terms are given by
\begin{align}
{\cal L}_{H} &= -m_{H_u}^2 H_u^{\dagger} H_u  -  m_{H_d}^2 H_d^{\dagger} H_d
         -  m_{\Phi_u}^2 \Phi_u^{\dagger} \Phi_u  -  m_{\Phi_d}^2 \Phi_d^{\dagger} \Phi_d \nonumber
\\
        &- m_{\Omega^+}^2 \Omega^{+ \, \dagger} \Omega^+  -  m_{\Omega^-}^2 \Omega^{- \, \dagger} \Omega^-
         -  m_{\zeta}^2 \zeta^{\dagger} \zeta  -  m_{\eta}^2 \eta^{\dagger} \eta \nonumber
\\
&- 
\left\{B\mu H_u H_d  +  B_{\Phi}\mu_{\Phi} \Phi_u \Phi_d  +  B_{\Omega}\mu_{\Omega} ( \Omega^+ \Omega^- + \zeta \eta) 
+\text{h.c.}\right\}\nonumber
\\
&- \left\{
A_{\zeta} H_d\Phi_u\zeta  +  A_{\eta} H_u\Phi_d\eta 
 +  A_{\Omega^-} H_u\Phi_u\Omega^-  +  A_{\Omega^+} H_d\Phi_d\Omega^+ 
+\text{h.c.}
\right\}\nonumber\\
&
-\left\{m_{\zeta\eta}^2\eta^{\dagger}\zeta
	+\frac{B_{\zeta}^2}{2}\zeta^2+\frac{B_{\eta}^2}{2}\eta^2
+\text{h.c.}\right\}
\;.
\label{eq:softbrreakingterms}
\end{align}

The coupling constant $\hat{\lambda}$ and the cutoff scale $\Lambda_H$ are related through the NDA.
Under the assumption of the NDA, the coupling constant $\hat{\lambda}$ becomes non-perturbative at $\Lambda_H$ as $\hat{\lambda}\simeq 4\pi$.
The value of $\hat{\lambda}$ at the cutoff scale $\Lambda_H$ is connected to 
those at low energy scales by RGE.
Therefore, the cutoff scale $\Lambda_H$ can be predicted from the value of the coupling constant
$\hat{\lambda}$ at the electroweak scale, $\hat{\lambda}(\mu_{\text{EW}})$.
In this paper, we constrain the range of $\Lambda_H$ by requiring that 
the coupling constant $\hat{\lambda}(\mu_{\text{EW}})$ satisfies
the condition of strongly 1stOPT, $\varphi_c/T_c\gtrsim 1$, which is one of the conditions
for successful electroweak baryogenesis\cite{EWBG}.
In general, non-decoupling quantum effects of additional scalar fields make 
the order of electroweak phase transition strong.
In Refs.~\cite{KSS, KSSY}, some of the extra scalar fields such as $\Phi_u$, $\Phi_d$, $\Omega^+$, $\Omega^-$
$\zeta$, and $\eta$ significantly contribute to make the order of electroweak phase transition stronger,
when the coupling $\hat{\lambda}$ satisfies $\hat{\lambda}(\mu_{\text{EW}}) \gtrsim 1.6$ at the electroweak scale.
Correspondingly, the Landau pole appears at the scale around ten TeV.

\section{Loop induced neutrino masses}

We will show that radiative seesaw scenarios\cite{aks,ma} are realized in the low energy effective 
theory of the $\text{SU}(2)_H\times Z_2$ model
by adding $Z_2$-odd RH neutrino superfields $N^c_i$.
The superpotential relevant to the neutrino sector is given by
\begin{align}
	W_{N}=&y_N^{ij}N^c_iL_j\Phi_u^{}+h_N^{ij}N^c_iE^c_j\Omega^-+\frac{M_i}{2}N^c_iN^c_i\;,
\end{align}
where $E^c_i$ and $L_i$ are the RH charged lepton chiral superfields and 
the lepton doublet chiral superfield, respectively, and 
the basis of the lepton fields are taken such that both the mass matrix for the $N^c_i$ and 
the charged lepton Yukawa matrix are real and diagonal.
Notice that the $Z_2$ parity prohibits the neutrino Yukawa interactions as $N^c_iL_jH_u$ which
give neutrino masses at the tree level, so that the type I seesaw mechanism does not work.

In our model, the neutrino masses are radiatively generated by 
(I) one-loop diagrams and (II) three-loop diagrams.
The one-loop diagrams correspond to the coupling constants $y_N^{ij}$, and 
the three-loop diagrams correspond to the coupling constants $h_N^{ij}$.

\subsection{One-loop contributions}

\begin{figure}
\begin{center}
	\includegraphics[scale=0.5]{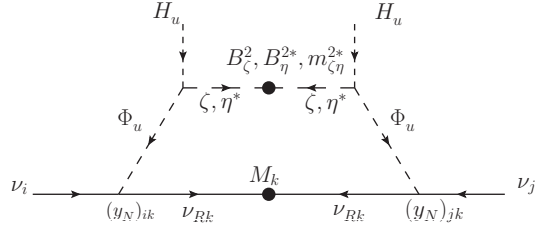}
\end{center}
\caption{A one-loop diagram which contributes to the neutrino mass matrix.}
\label{Fig:1loop}
\end{figure}

The one-loop diagrams which contribute to the neutrino mass matrix are shown in 
Fig.~\ref{Fig:1loop}.
These diagrams correspond to the SUSY extension of the Ma model\cite{ma}.
Such mass terms as $\eta^2$ or $\zeta^2$ cannot be written in the superpotential 
of our model due to the SUSY dynamics at $\Lambda_H$, so that
the loop diagrams with RH sneutrinos and $Z_2$-odd fermions
do not contribute.
The contributions to the mass matrix are calculated as 
\begin{align}
	m_{ij}^{(\text{I})}=\frac{(y_N^{})^{ki}(y_N^{})^{kj}}{(4\pi)^2}
	\left\{
		(O_0^{})^{1\alpha}(O_0^{})^{1\alpha}M_{k}^{}
	-(O_0^{})^{5\alpha}(O_0^{})^{5\alpha}M_{k}^{}\right\}\bar{B}_0(m_{\Phi_{\alpha}}^{2},M_{k}^{2})\;,
\end{align}
where the loop function $\bar{B}_0$ is given as
\begin{equation}
	\bar{B}_0^{}(m_1^2,m_2^2)=
	-\frac{m_1^2\ln m_1^2-m_2^2\ln m_2^2}{m_1^2-m_2^2}\;,
\end{equation}
and the matrix $O_0$ is the mixing matrix for the $Z_2$-odd neutral scalars (see Appendix~A).

\subsection{Three-loop contributions}
\begin{figure}
\begin{center}
	\begin{tabular}{cc}
		\includegraphics[scale=0.45]{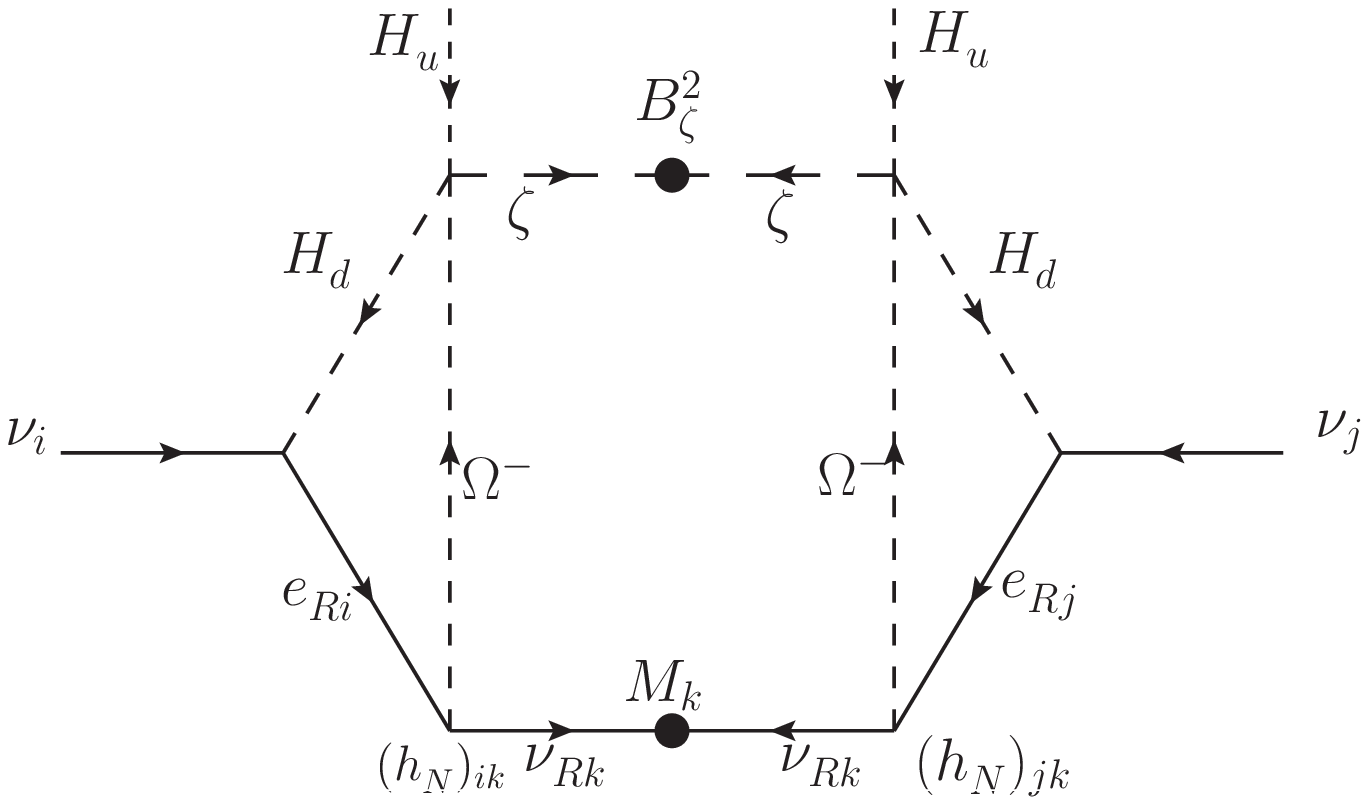}&
		\includegraphics[scale=0.45]{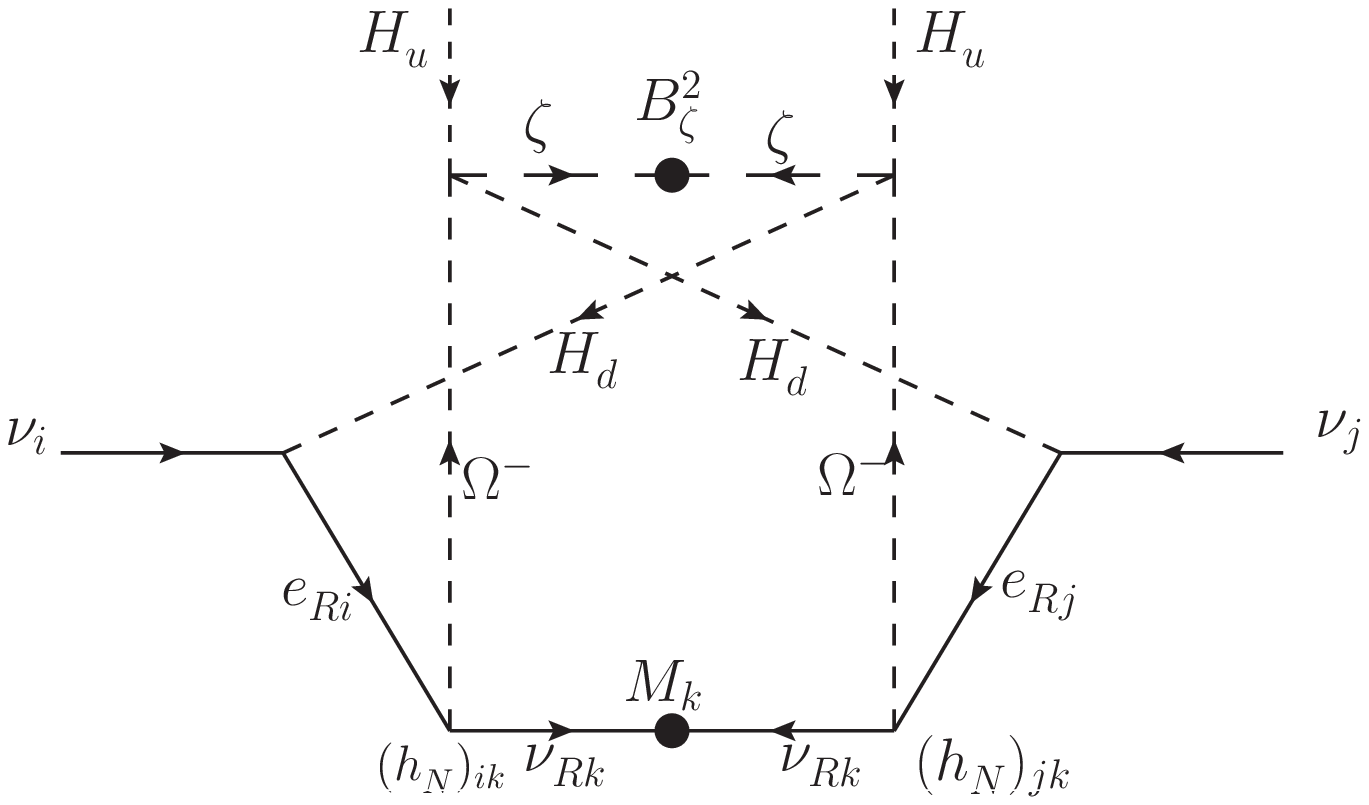}\\
		\includegraphics[scale=0.45]{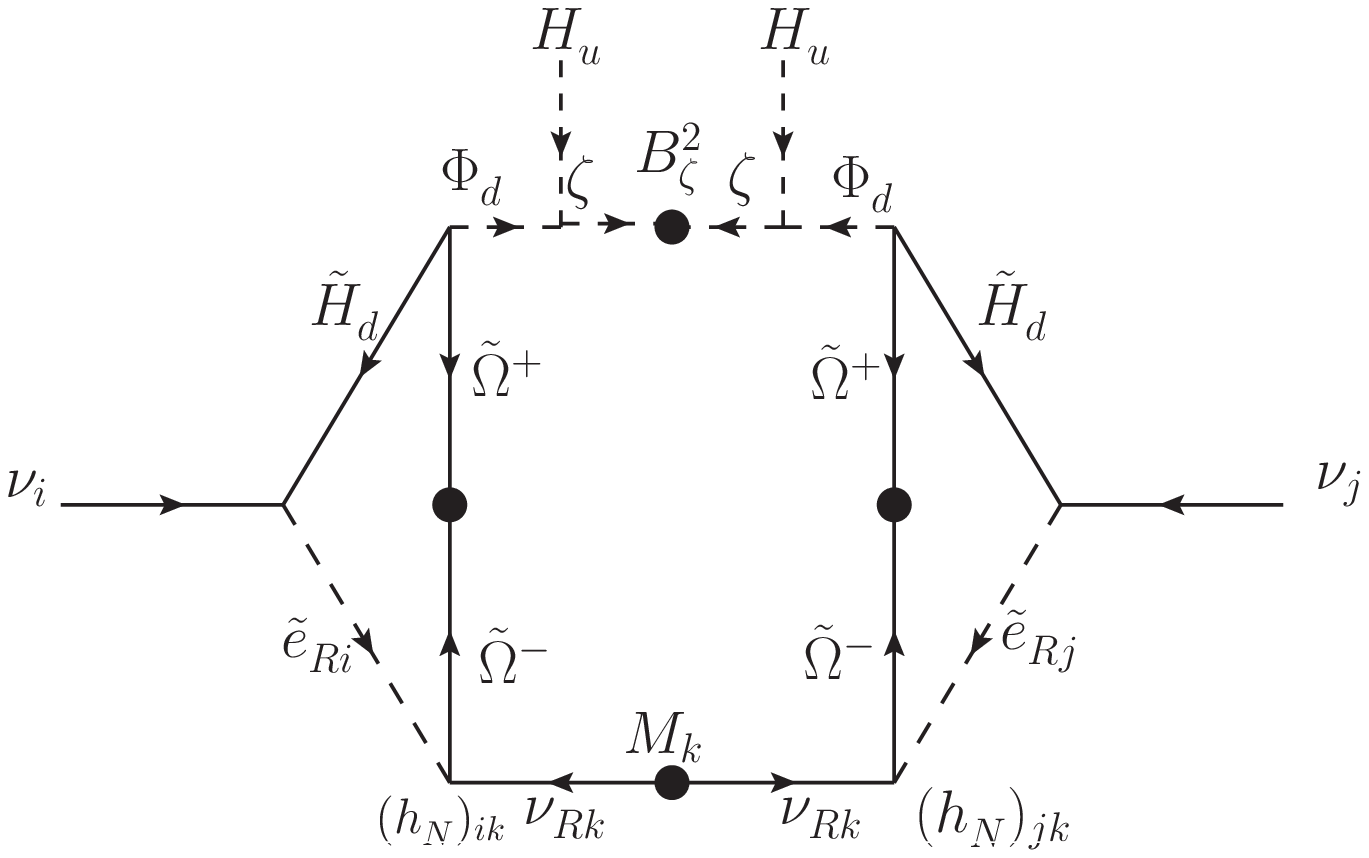}&
		\includegraphics[scale=0.45]{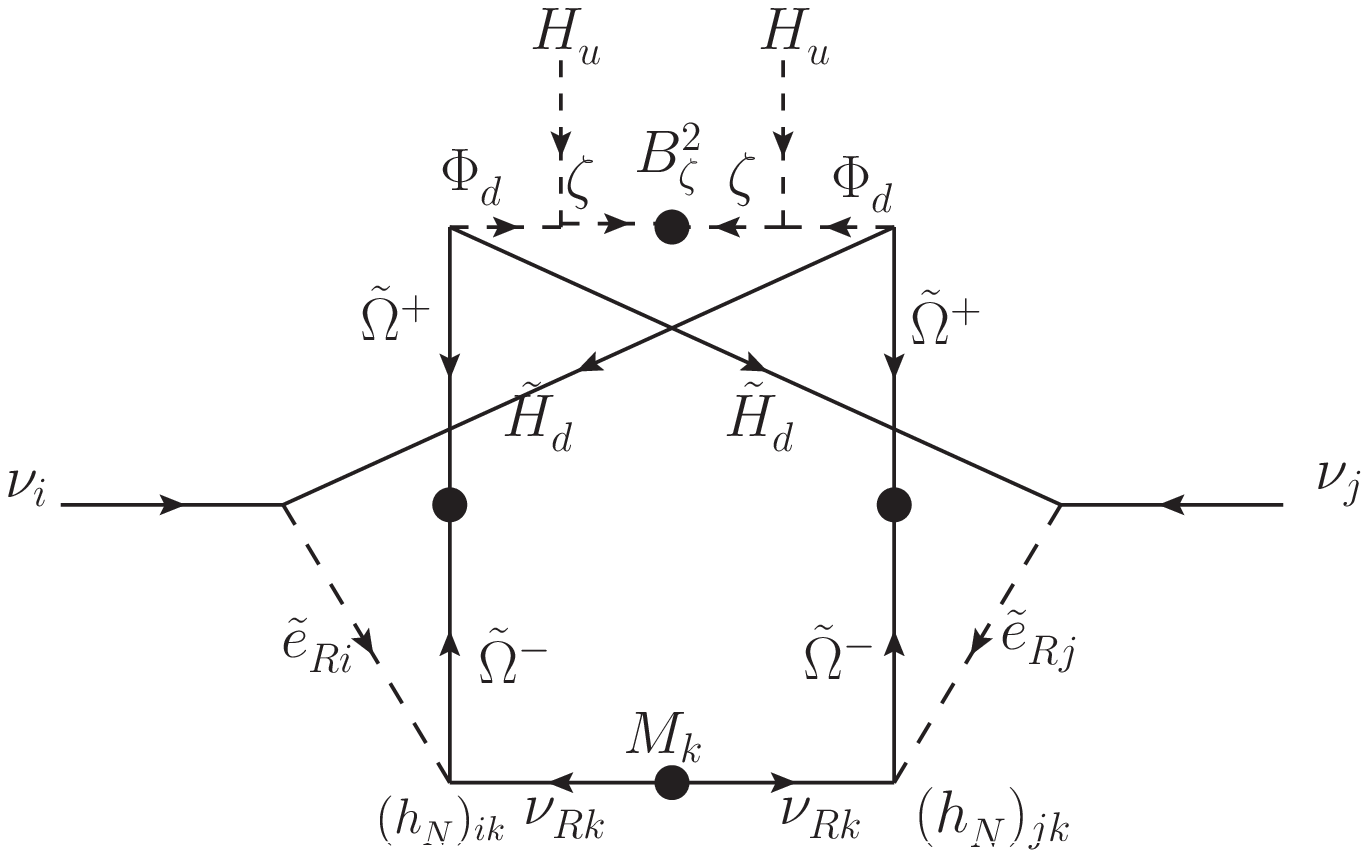}\\
\end{tabular}
\end{center}
\caption{Three-loop diagrams which contribute to the neutrino mass matrix.}
\label{Fig:3loop}
\end{figure}

The three-loop diagrams which contribute to the neutrino mass matrix are 
shown in Fig.~\ref{Fig:3loop}.
The contributions are calculated as
\begin{align}
	m_{ij}^{\text{(II)}}=
	\displaybreak[0]
	&\frac{\hat{\lambda}^4v_u^2(y_E)_i(h_N^{*})_{ki}(y_E)_j(h_N^{*})_{kj}M_k}{(16\pi^2)^3}\nonumber\\
	\displaybreak[0]
	&\times
	\sin^4\beta
	(U_{+}^*)_{4\gamma}(U_{+})_{4\gamma}
	(U_{+}^*)_{4\delta}(U_{+})_{4\delta}
	\left\{(O_0)_{2\rho}(O_0)_{2\rho}-(O_0)_{6\rho}(O_0)_{6\rho}\right\}
	\nonumber\\
	\displaybreak[0]
	&\times 
	F(M_k^2,m_{\Phi_{\rho}}^2;m_{e_i}^2,m_{H^{\pm}}^2,m_{\Phi^{\pm}_{\gamma}}^2;m_{e_j}^2,m_{H^{\pm}}^2,m_{\Phi^{\pm}_{\delta}}^2)
	\nonumber\\
	\displaybreak[0]
	&+\frac{2\hat{\lambda}^2(y_E)_i(h_N^{*})_{ki}(y_E)_j(h_N^{*})_{kj}M_km_{\tilde{\Phi}^{\pm}_{\gamma}}
m_{\tilde{\Phi}^{\pm}_{\delta}}}{(16\pi^2)^3}\nonumber\\
	\displaybreak[0]
	&\times
	(V_{L}^*)_{2\alpha}(V_{L})_{2\alpha}
	(V_{L}^*)_{2\beta}(V_{L})_{2\beta}
	({U}_L^*)_{2\gamma}({U}_R)_{2\gamma}
	({U}_L^*)_{2\delta}({U}_R)_{2\delta}
	\nonumber\\
	\displaybreak[0]
	&\times\left\{(O_0)_{3\rho}(O_0)_{3\rho}-(O_0)_{7\rho}(O_0)_{7\rho}\right\}
	F(M_k^2,m_{\Phi_{\rho}}^2;m_{\tilde{\chi}^{\pm}_{\alpha}}^2,m_{\tilde{e}_{Ri}}^2,m_{\tilde{\Phi}^{\pm}_{\gamma}}^2;
	m_{\tilde{\chi}^{\pm}_{\beta}}^2,m_{\tilde{e}_{Rj}}^2,m_{\tilde{\Phi}^{\pm}_{\delta}}^2)
	\;.
\end{align}
where the loop function $F$ is given by\cite{aks2}
\begin{align}
	&F(M^2,m_{\Phi}^2,m_{\chi_1}^2;m_{\phi_1}^2,m_{\Omega_1}^2;m_{\chi_2}^2,m_{\phi_2}^2,m_{\Omega_2}^2)\nonumber\\
	&=
	\frac{(16\pi^2)^3}{i}\int\frac{d^Dk}{(2\pi)^D}\frac{1}{k^2-M^2}\frac{1}{k^2-m_{\Phi}^2}
\int\frac{d^Dp}{(2\pi)^D}
\frac{\Fdagger{p}}{p^2-m_{\chi_1}^2}
\frac{1}{p^2-m_{\phi_1}^2}
\frac{1}{(k+p_1)^2-m_{\Omega_1}^2}\nonumber\\
&\phantom{SpaceSpace}\times 
\int\frac{d^Dq}{(2\pi)^D}
\frac{(-\Fdagger{q})}{(-q)^2-m_{\chi_2}^2}
\frac{1}{(-q)^2-m_{\phi_2}^2}
\frac{1}{(k+(-q))^2-m_{\Omega_2}^2}\nonumber\\
&=
\frac{1}{(M^2-m_{\Phi}^2)(m_{\chi_1}^2-m_{\phi_1}^2)(m_{\chi_2}^2-m_{\phi_2}^2)}
\int_0^{\infty}k_E^2d(k_E^2)\left(
	\frac{M^2}{-k_E^2-M^2}-\frac{m_{\Phi}^2}{-k_E^2-m_{\Phi}^2}
\right)
\nonumber\\
&\phantom{Spac}\times
\left\{
	\bar{B}_1(-k_E^2,m_{\chi_1}^2,m_{\Omega_1}^2)-\bar{B}_1(-k_E^2,m_{\phi_1}^2,m_{\Omega_1}^2)
\right\}\nonumber\\
&\phantom{Spac}\times
\left\{
	\bar{B}_1(-k_E^2,m_{\chi_2}^2,m_{\Omega_2}^2)-\bar{B}_1(-k_E^2,m_{\phi_2}^2,m_{\Omega_2}^2)
\right\}
\;,
\end{align}
with $\bar{B}_1$ being
\begin{equation}
	\bar{B}_1(p^2,m_1^2,m_2^2)\equiv 
	-\int_0^1dx x\ln\frac{(1-x)m_1^2+xm_2^2-x(1-x)p^2-i\varepsilon}{\mu^2}\;.
\end{equation}
The numerical behavior of the improper integrals in evaluation of the function $F$ 
is discussed in Ref.~\cite{aks2}.
The matrices $U_+$, $U_L$ and $U_R$ are  
mixing matrices for $Z_2$-odd charged particles as given 
in Appendix~A,
while the matrices ${V}_L$ and ${V}_R$ are the mixing matrices for 
the MSSM charginos as 
\begin{equation}
	V_R^{\dagger}
	\begin{pmatrix}
		M_{\tilde{W}}&\sqrt{2}m_W\cos\beta\\
		\sqrt{2}m_W\sin\beta&\mu
	\end{pmatrix}
	V_L=
	\begin{pmatrix}
		m_{\tilde{\chi}_1}^{}&0\\
		0&m_{\tilde{\chi}_2}^{}
	\end{pmatrix}
	\;,
\end{equation}
where $M_{\tilde{W}}$ is the wino mass.
This is a SUSY extension of the AKS model\cite{aks,aks2}\footnote{
In the original non-SUSY AKS model, the Higgs sector is the type-X 
two Higgs doublet model with neutral and charged singlet fields.
The type-X two Higgs doublet model is adopted in order to make 
the charged Higgs boson light with avoiding too large contribution 
to the $b\to s \gamma$ process.
On the other hand, in the model discussed here, the $Z_2$-even Higgs 
sector is the type II two Higgs doublet model and the constraint from 
$b\to s\gamma$ can be satisfied with the charged Higgs mass taken in the 
benchmark points.
In spite of such a small difference, one can say that the model is essentially 
identical to the SUSY extended AKS model.
}.
In the AKS model, extra neutral and charged singlet scalar fields are added to
a two Higgs doublet model. 
The chiral superfields $\zeta$ and $\Omega^-$ correspond to these extra singlet 
scalar fields.
In the SUSY extended AKS model, 
an extra doublet superfield $\Phi_d$ is necessary to 
provide an indispensable quartic scalar interaction such as 
$H_uH_d^{\dagger}\Omega^-\zeta^*$  by F-term.
The superfields $\Phi_u$ and $\Omega^+$ are required for chiral anomaly cancellation.
It is surprising that all the superfields required in the SUSY AKS model are 
automatically provided in the $\text{SU(2)}_H\times Z_2$ model.

\subsection{Benchmark points}
We here consider the benchmark points where 
the neutrino oscillation data can be reproduced 
in addition to make 1stOPT
strong as $\varphi_c/T_c\gtrsim 1$ in the $\text{SU(2)}_H\times Z_2$ model.
In general, both the one-loop and the three-loop diagrams contribute to the neutrino mass generation.
However, we here consider the following two limiting cases: (A) one-loop dominant case ($h_N^{ij}=0$), 
and (B) three-loop dominant case ($y_N^{ij}=0$).
The definition of the two benchmark points are shown in Table~\ref{Tab:Bench}.
The mass of the SM-like Higgs boson is tuned to be $m_h=125$\;GeV by 
choosing the parameters in the scalar top sector; {\it i.e.}, SUSY breaking soft masses and 
left-right mixing parameter of the stops.
For simplicity, we do not put any additional flavor mixing in the scalar lepton mass matrices.
\begin{table}
	\caption{Benchmark parameter set for (A) the one-loop dominant case and (B) 
	three-loop dominant case. 
	For both cases, $B_{\Phi}=B_{\Omega}=A_{\zeta}=A_{\eta}=A_{\Omega^+}=
	A_{\Omega^-}=0$ is taken.\label{Tab:Bench}}
	\begin{center}
\begin{tabular}{l}
	\begin{tabular}{|c||c|c|c|c|c|c|c|}\hline
		Case &$\hat{\lambda}$	&$\tan\beta$&$m_{H^{\pm}}$&$m_{\tilde{W}}$&$\mu$
		&$\mu_{\Phi}$&$\mu_{\Omega}$
		\\ \hline
		(A) &$1.8$		&$15$		& $350$\;GeV &$500$\;GeV&$100$\;GeV&$550$\;GeV&$-550$\;GeV
		\\ \hline
		(B) &$1.8$		&$30$		& $350$\;GeV &$500$\;GeV&$100$\;GeV&$550$\;GeV&$-550$\;GeV
		\\ \hline
	\end{tabular}\\[10mm]
	\begin{tabular}{|c||c|c|c|c|c|c|} \hline
		Case
			&$\bar{m}_{\Phi_u}^2$
			&$\bar{m}_{\Phi_d}^2$
			&$\bar{m}_{\Omega^+}^2$
			&$\bar{m}_{\Omega^-}^2$
			&$\bar{m}_{\zeta}^2$
			&$\bar{m}_{\eta}^2$ \\ \hline
		(A)
		&$(100\;\text{GeV})^2$
		&$(1500\;\text{GeV})^2$
		&$(1500\;\text{GeV})^2$
		&$(100\;\text{GeV})^2$
		&$(1500\;\text{GeV})^2$
		&$(2000\;\text{GeV})^2$
		\\ \hline
		(B)
		&$(1500\;\text{GeV})^2$
		&$(1500\;\text{GeV})^2$
		&$(1500\;\text{GeV})^2$
		&$(30\;\text{GeV})^2$
		&$(1410\;\text{GeV})^2$
		&$(30\;\text{GeV})^2$
		\\ \hline
	\end{tabular}\\[10mm]
	\begin{tabular}{|c||c|c|c|c|c|c|} \hline
		Case&$B_{\zeta}^2$&$B_{\eta}^2$&$m_{\zeta\eta}^2$
		\\ \hline
		(A)
		&$(100\;\text{GeV})^2$
		&$(100\;\text{GeV})^2$
		&$(100\;\text{GeV})^2$
		\\ \hline
		(B)
		&$(1400\;\text{GeV})^2$
		&$0$
		&$0$
		\\ \hline
	\end{tabular}\\[10mm]
	\begin{tabular}{|c||c|c|c|c|c|c|c|} \hline
		Case&$M_1$&$M_2$&$M_3$
		&$m_{\tilde{\nu}_{R1}}$
		&$m_{\tilde{\nu}_{R2}}$
		&$m_{\tilde{\nu}_{R3}}$
		&$m_{\tilde{e}_{Ri}}(i=1,2,3)$
		\\ \hline
		(A)
		&$60$\;GeV
		&$120$\;GeV
		&$180$\;GeV
		&$60$\;GeV
		&$120$\;GeV
		&$180$\;GeV
		&$5000$\;GeV
		\\ \hline
		(B)
		&$100$\;GeV
		&$2000$\;GeV
		&$4000$\;GeV
		&$100$\;GeV
		&$3000$\;GeV
		&$5000$\;GeV
		&$5000$\;GeV
		\\ \hline
	\end{tabular}\\[10mm]
	\begin{tabular}{|c||c|c|}\hline
		Case&$(y_N)_{ij}$&$(h_N)_{ij}$\\\hline
			  &&\\[-5mm]
		(A)
		&$\begin{pmatrix}
	-0.439& -0.424&0.512\\
	 0.226&0.218& -0.263\\ 
	 0.272&1.36&1.36
	\end{pmatrix}\times 10^{-4}$
	&
	$\begin{pmatrix}
	0&0&0\\
	0&0&0\\
	0&0&0\\
\end{pmatrix}$
\\[10mm] \hline
			  &&\\[-5mm]
		(B)
		&
		$\begin{pmatrix}
			0&0&0\\
			0&0&0\\
			0&0&0\\
		\end{pmatrix}$
		&
		$\begin{pmatrix}
			0.003&0&0\\
  	-0.0164 - 1.26 i& -0.02424 + 0.0049 i& -0.0022 + 0.00097 i\\
	0.491 - 1.581 i& 0.02461 + 0.00537 i& 0.0016 + 0.0019 i
		\end{pmatrix}$
	\\[10mm] \hline
	\end{tabular}
\end{tabular}
\end{center}
\end{table}

We will discuss consequences of the benchmark points.
First, we will show the strength of 1stOPT
$\varphi_c/T_c$ and related issues in the Table~\ref{Tab:EWPT}.
In order to satisfy $\varphi_c/T_c>1$ by the mechanism discussed in Ref.~\cite{KSS,KSSY}, 
we take $\hat{\lambda}=1.8$ which leads to 
the cut-off scale at around $\Lambda_H=5$\;TeV on both benchmark points.
The enhancement occurs by the non-decoupling loop contributions of $Z_2$-odd scalars.
These non-decoupling loop contributions affect the triple coupling of the SM-like Higgs boson 
$\lambda_{hhh}$, and loop effects of $Z_2$-odd charged scalars can deviate the decay 
branching ratio of the Higgs boson into diphoton $\text{B}(h\to \gamma\gamma)$ from 
the SM prediction.
The ratio of $\lambda_{hhh}$ to its SM prediction and 
the ratio of $\text{B}(h\to \gamma\gamma)$ to its SM prediction are
evaluated for each of the benchmark points as shown in Table~\ref{Tab:EWPT}, and
one find 10-20\% deviations for them.
\begin{table}[ht]
	\caption{
		The predicted value of the cut-off scale $\Lambda_H$, $\varphi_c/T_c$, 
		the ratio of the coupling constant $\lambda_{hhh}$ to its SM prediction $\lambda_{hhh}^{}/\left.\lambda_{hhh}^{}\right|_{\text{SM}}$,
		and the ratio of the branching ratio $\text{B}(h\to \gamma\gamma)$ to its SM prediction 
		$\text{B}(h\to \gamma\gamma)/\left.\text{B}(h\to \gamma\gamma)\right|_{\text{SM}}$.
		\label{Tab:EWPT}
	}
	\begin{tabular}{|c||c|c|c|c|}\hline
		Case&$\Lambda_H$&$\varphi_c/T_c$&$\lambda_{hhh}^{}/\left.\lambda_{hhh}\right|_{\text{SM}}$&$\text{B}(h\to \gamma\gamma)/\left.\text{B}(h\to \gamma\gamma)\right|_{\text{SM}}$\\ \hline
		(A)&$5$\;TeV&$1.0$&$1.18$&$0.80$\\ \hline
		(B)&$5$\;TeV&$1.2$&$1.09$&$0.89$\\ \hline
	\end{tabular}
\end{table}

To see the detail of the non-decoupling effects on the condition of $\varphi_c/T_c$, 
$\lambda_{hhh}$ and $\text{B}(h\to \gamma\gamma)$, we show the 
mass spectrum of $Z_2$-odd particles in Table~\ref{Tab:Z2oddmass}.
In the case (A), the spectrum is very similar to the one given in Ref.~\cite{KSSY}.
There, the charged scalar eigenstate $\Phi_1^{\pm}$ and $\Phi_2^{\pm}$ are almost 
from the charged scalar components of $\Omega^-$ and $\Phi_u$ respectively, and 
their masses are dominated by the $\hat{\lambda}^2v^2$ terms.
So significant non-decoupling effects appear in 1stOPT,
$\lambda_{hhh}$ and $\text{B}(h\to \gamma\gamma)$.
In the neutral $Z_2$-odd scalar sector, there is no significant non-decoupling effects, 
because all the mass eigenvalues are not dominated by the Higgs vev contributions.
On the other hand, in the case (B), the eigenstates $\Phi_2^0$ and $\Phi_3^0$ 
which are almost from the neutral components of $\eta$ give significant contributions 
to $\varphi_c/T_c$ and $\lambda_{hhh}$.
In addition, the non-decoupling effect by $\Phi_1^{\pm}\sim \Omega^-$ 
contributes to $\varphi_c/T_c$, $\lambda_{hhh}$ and $\text{B}(h\to \gamma\gamma)$ 
as same as in the case (A).
\begin{table}[ht]
	\caption{
		The mass spectrum for the $Z_2$-odd particles obtained from 
		the benchmark points defined in Table~\ref{Tab:Bench}.
		\label{Tab:Z2oddmass}
	}
	\begin{tabular}{l}
	\begin{tabular}{|c||c|c|c|c|c|c|c|c|}\hline
		Case& \multicolumn{8}{c|}{$Z_2$-odd neutral bosons}\\ \cline{2-9}
		   & $\Phi_1^0$& $\Phi_2^0$& $\Phi_3^0$& $\Phi_4^0$& $\Phi_5^0$& $\Phi_6^0$
		   & $\Phi_7^0$& $\Phi_8^0$ \\ \hline
		(A)
		& $88.3$\;GeV
		& $88.5$\;GeV
		& $1457$\;GeV
		& $1462$\;GeV
		& $1569$\;GeV
		& $1571$\;GeV
		& $2023$\;GeV
		& $2028$\;GeV
		\\ \hline
		(B)
		& $126$\;GeV
		& $294$\;GeV
		& $294$\;GeV
		& $1505$\;GeV
		& $1506$\;GeV
		& $1525$\;GeV
		& $1535$\;GeV
		& $1992$\;GeV
		\\ \hline
	\end{tabular}\\[13mm]
	\begin{tabular}{|c||c|c|c|c|}\hline
		Case& \multicolumn{4}{c|}{$Z_2$-odd charged bosons}\\ \cline{2-5}
		   & $\Phi_1^{\pm}$& $\Phi_2^{\pm}$& $\Phi_3^{\pm}$& $\Phi_4^{\pm}$ \\ \hline
		(A)
		& $288$\;GeV
		& $307$\;GeV
		& $1496$\;GeV
		& $1517$\;GeV
		\\ \hline
		(B)
		& $271$\;GeV
		& $1459$\;GeV
		& $1506$\;GeV
		& $1574$\;GeV
		\\ \hline
	\end{tabular}\\[13mm]
	\begin{tabular}{|c||c|c|c|c|}\hline
		Case& \multicolumn{4}{c|}{$Z_2$-odd neutral fermions}\\ \cline{2-5}
		& $\tilde{\Phi}_1^{0}$& 
		$\tilde{\Phi}_2^{0}$& $\tilde{\Phi}_3^{0}$& $\tilde{\Phi}_4^{0}$ \\ \hline
		(A)
		& $429$\;GeV
		& $429$\;GeV
		& $721$\;GeV
		& $721$\;GeV
		\\ \hline
		(B)
		& $422$\;GeV
		& $422$\;GeV
		& $725$\;GeV
		& $725$\;GeV
		\\ \hline
	\end{tabular}\\[13mm]
	\begin{tabular}{|c||c|c|}\hline
		Case& \multicolumn{2}{c|}{$Z_2$-odd charged fermions}\\ \cline{2-3}
		& $\tilde{\Phi}_1^{\pm}$& 
		$\tilde{\Phi}_2^{\pm}$ \\ \hline
		(A)
		&\phantom{S}$429$\;GeV\phantom{S}
		&$721$\;GeV
		\\ \hline
		(B)
		&$422$\;GeV
		&$725$\;GeV
		\\ \hline
	\end{tabular}
\end{tabular}
\end{table}

Next, we will show the neutrino masses and mixing angles obtained on the benchmark 
points.
In order to obtain the neutrino mass scale of order of 0.1\;eV, the constants $y_N^{ij}$ 
in the case of (A) are $\mathcal{O}(10^{-4})$. 
On the other hand, in the case of (B), some elements of $h_N^{ij}$ are required to be 
rather large.
Especially, in order to compensate the suppression by the small electron Yukawa coupling, 
the magnitudes of couplings $h_N^{1i}$ are of order one.
With the coupling constant matrices $y_N^{ij}$ and $h_N^{ij}$ given in Table~\ref{Tab:Bench},
the neutrino mass eigenvalues and the mixing angles are obtained as displayed in 
Table~\ref{Tab:neutrino}.
These predicted values are in the allowed region which is given 
by the global fitting analysis of neutrino oscillation data as\cite{nufit}
\begin{align}
&2.28<\frac{|m_3^2-m_1^2|}{10^{-3}\;\text{eV}^2}<2.70\;,\quad
7.0<\frac{m_2^2-m_1^2}{10^{-5}\;\text{eV}^2}<8.1,\nonumber\\
&0.27<\sin^2\theta_{12}<0.34\;,\quad
0.34<\sin^2\theta_{23}<0.67\;,\quad
0.016<\sin^2\theta_{13}<0.030\;,
\end{align}
where $m_i (i=1,2,3)$ are the mass eigenvalues of the 
neutrinos, and $\theta_{12}$, $\theta_{23}$, and $\theta_{13}$
are the mixing angles relevant to 
the solar neutrino mixing, atmospheric neutrino mixing and 
the reactor neutrino mixing respectively.

\begin{table}[ht]
	\caption{
		The neutrino masses and mixing angles obtained on
		the benchmark points defined in Table~\ref{Tab:Bench}.
			\label{Tab:neutrino}
	}
	\begin{tabular}{|c||c|c|c|c|c|c|c|c|} \hline
		Case
		&$m_1$
		&$m_2$
		&$m_3$
		&$\sin^2\theta_{12}$
		&$\sin^2\theta_{23}$
		&$|\sin\theta_{13}|$
		\\ \hline
		(A)
		&$0.0\;\text{eV}$
		&$0.0087\;\text{eV}$
		&$0.050\;\text{eV}$
		&$0.31$
		&$0.50$
		&$0.14$
		\\ \hline
		(B)
		&$0.0\;\text{eV}$
		&$0.0084\;\text{eV}$
		&$0.050\;\text{eV}$
		&$0.32$
		&$0.50$
		&$0.14$
		\\ \hline
	\end{tabular}
\end{table}

The coupling constants $y_N^{ij}$ and $h_N^{ji}$ can give significant contributions 
to some of the lepton flavor violation processes through 
the RH neutrino and sneutrino mediation diagrams.
The predicted values of the branching ratios $\text{B}(\mu\to e\gamma)$ and 
$\text{B}(\mu\to eee)$ are listed in Table~\ref{Tab:LFV}.
In the case (A), as already discussed, the coupling constants $y_N^{ij}$ are so small 
that the contribution to the $\mu\to e\gamma$ is suppressed enough 
to satisfy the current upper bound given by the MEG experiment
$\text{B}(\mu\to e\gamma)\leq 5.7\times 10^{-13}$\cite{MEG}.
In addition, the branching ratio of the $\mu\to eee$ is approximately 
given as 
\begin{equation}
	\text{B}(\mu\to eee)\sim \frac{\alpha}{4\pi}\text{B}(\mu\to e\gamma)\;.
\end{equation}
Then the experimental upper bound on the branching ratio such as
$\text{B}(\mu\to eee)\leq 10^{-12}$\cite{meee} 
is satisfied once the $\mu\to e\gamma$ is suppressed enough.
In the case (B), on the other hand, 
large coupling constants $h_N^{1i}$ enhance the $\mu\to e\gamma$ process.
The constraint from $\text{B}(\mu\to eee)$ is also severe in this case,
even if the branching ratio $\text{B}(\mu\to e\gamma)$ is suppressed enough\cite{aky}.
It is because the order one coupling constants $h_N^{1i}$ enhance the contributions 
from box diagram where the RH neutrinos and RH sneutrinos are running in the loop.
The predicted values of $\text{B}(\mu\to e\gamma)$ and $\text{B}(\mu\to eee)$ 
on the benchmark points are shown in Table~\ref{Tab:LFV}, and we find that they satisfy 
these experimental upper bounds on both benchmark points.
In the case (B), since the branching ratio $\text{B}(\mu\to e\gamma)$ is predicted just 
below the current limit, it is expected that the $\mu\to e\gamma$ process will be observed 
in future experiments.
\begin{table}
	\caption{
		The prediction on the branching ratios of lepton flavor violation processes
		$\text{B}(\mu\to e\gamma)$ and $\text{B}(\mu\to eee)$ on
		the benchmark points defined in Table~\ref{Tab:Bench}.
		\label{Tab:LFV}
	}
	\begin{tabular}{|c||c|c|} \hline
		Case
		&$\text{B}(\mu\to e\gamma)$
		&$\text{B}(\mu\to eee)$
		\\ \hline
		(A)
		&$5.2\times 10^{-19}$
		&$8.1\times 10^{-21}$
		\\ \hline
		(B)
		&$5.0\times 10^{-13}$
		&$8.5\times 10^{-13}$
		\\ \hline
	\end{tabular}
\end{table}

We have found that the benchmark points defined in 
Table~\ref{Tab:Bench} can reproduce
the correct values of neutrino masses and mixing angles with satisfying the 
constraint from lepton flavor violations and with keeping strong enough 1stOPT for 
electroweak baryogenesis.

\subsection{Collider signatures}
In this paper, we do not perform any complete analysis of 
specific collider signals.
We here give some comments, and 
detailed analysis of collider signatures in our model will be 
discussed elsewhere.

\subsubsection{Precise measurements of the Higgs couplings}
As shown in Ref.~\cite{KSSY}, in the parameter region where 
1stOPT becomes strong enough 
for successful electroweak baryogenesis, 
the non-decoupling effect gives significant contributions to 
Higgs couplings such as the $hhh$ coupling and the $h\gamma\gamma$ coupling.
The direction of deviations for these coupling constants are related to each 
other.
Both couplings can deviate as large as $20$\% from the SM 
predictions, which can be tested by future collider experiments.
At the LHC, the branching ratio of Higgs to diphoton process 
will be measured at about 20\% accuracy, but the measurement 
of triple Higgs boson coupling is very challenging.
At the HL-LHC with the luminosity of 3000\;$\text{fb}^{-1}$, 
$\text{B}(h\to \gamma\gamma)$ will be measured with 10\% accuracy\cite{HLLHC}. 
The triple Higgs boson coupling can be measured at the HL-LHC 
and much better at the ILC.  It is expected that the $hhh$ coupling 
can be measured with the accuracy of about 20\;\% or better 
at the ILC with $\sqrt{s}=1$\;TeV with 2\;ab$^{-1}$\cite{TDRILC}.

\subsubsection{Direct search of the extra particles}
There are many extra fields which can provide collider signals in our model.
The $Z_2$-even sector of our model is essentially same as 
the nMSSM. Therefore we can expect that the collider signals relevant to 
the $Z_2$-even particles are same as them in the nMSSM which are studied 
in the literature\cite{collidernMSSM}.

Our model is characterized by the $Z_2$-odd sector, so that
the collider signals in this sector are very important.
In the case (A) of our benchmark points, inert doublet-like scalars are 
light.
Collider signatures of the inert doublet scalars have been studied 
in the literature\cite{ColliderID,AokiKanemura, AKYok}. The inert doublet scalars are 
color singlet particles, then it is not easy to discover them 
at the LHC. Even though they can be fortunately discovered at the LHC, 
precise determination of their masses and quantum numbers are challenging\cite{ColliderID}.
On the other hand, the ILC is a very powerful tool to study such non-colored inert doublet particles.
At the ILC, the mass of charged inert scalar can be measured 
in a few GeV accuracy, and 
the mass of neutral inert scalar can be measured in better than 2\;GeV 
accuracy\cite{AKYok}.

In the case (B), the $Z_2$-odd singlet-like charged particle is required to be light.
As discussed in Ref.~\cite{aks2}, such the light singlet-like charged particle 
can be studied at the ILC via the pair production such as $e^+e^-\to \varphi_1^+\varphi_1^-$.
Furthermore, due to the interaction of $N_i^cE_j^c\Omega^-$, 
the production process such as  $e^-e^-\to \varphi_1^-\varphi_1^-$ is possible.
This process will be a strong evidence of three-loop neutrino mass generation 
mechanism\cite{AokiKanemura}.
The process can be detected at the $e^-e^-$ collision option of the ILC or 
the CLIC\cite{aks2,AokiKanemura}.

In addition, the SUSY extended Higgs sector of this model includes several 
color singlet SUSY partner fermions of the extra scalars.
If such SUSY partner particles are discovered, it discriminates
our model from non-SUSY models with radiative seesaw scenarios.

\subsection{Discussions}

\subsubsection{Evaluation of the baryon asymmetry of the Universe}
For baryogenesis, we focus on the strength of 1stOPT
which gives a necessary condition for successful electroweak 
baryogenesis, and we have not numerically evaluated the prediction on the BAU
in our scenario. 
In order to complete the numerical evaluation of the BAU,
we should also take care of the CP phases.
Since it is known that the CP violation in the SM is too small for getting 
the enough large BAU\cite{CPinSM},
extra CP phases are also required in addition to the mechanism 
to enhance 1stOPT.
In SUSY models, new sources of the CP violation which can contribute
to the generation of the BAU can be introduced\cite{CPinSUSY}.
In the literature\cite{SUSYEWBG}, numerical evaluation of the BAU
due to the electroweak baryogenesis in the MSSM is discussed.
In principle, we can introduce the CP phases to the model in the similar ways 
as the works mentioned above.
We then expect to obtain sufficient amount of BAU,
once the strong enough 1stOPT is realized.

\subsubsection{Dark matter} 
This model includes an unbroken $Z_2$ parity, which provides DM candidates.
Since the $Z_2$-odd extra fields except for the RH neutrino have quite strong
coupling with the Higgs bosons, they conflict with the bounds from direct 
detection experiments of the DM. 
We choose the both benchmark points in such a way that 
the lightest $Z_2$-odd particle is the RH neutrino and/or the RH sneutrino.
If the $R$ parity is also imposed, the lightest SUSY particle also qualifies 
as the DM candidate.
It leads to a rich possibility of the multi-component DM scenario\cite{multiDM}.
In this paper we do not specify the scenario of DM.
Detailed analysis of the relic abundance and the direct detection constraints
are performed elsewhere.

\subsubsection{Mediation mechanism of the SUSY breaking}
Due to the non-renormalization theorem, the neutrino masses are 
not generated supersymmetric; {\it i.e.}, soft SUSY breaking 
terms are necessary for loop induced neutrino mass models.
In our model, SUSY breaking terms in the last line of Eq.~(\ref{eq:softbrreakingterms})
are essential. 
These terms are not forbidden by the gauge symmetry, but 
no relevant terms are in the superpotential given in Eq.~(\ref{eq:superpotentialWeff}).
It may suggest a specific mediation mechanism for the SUSY breaking.
It is a quite interesting point that the neutrino mass generation is a key to 
explore the mediation mechanism of SUSY breaking in our model.

\section{Conclusions}

We have considered a model based on the SUSY gauge theory with 
$N_c=2$ and $N_f=3$ with an additional exact $Z_2$ symmetry.
By adding $Z_2$-odd RH neutrinos to the model, 
we have proposed a concrete model which can be a fundamental theory 
of a low energy effective theory with radiative seesaw scenarios and 
with strong 1stOPT.
We have shown that radiative seesaw scenarios can be realized 
in our model and there can be two types of contributions to 
the neutrino mass matrix;
{\it i.e.}, by one-loop diagrams and also by three-loop diagrams.
These contributions correspond to the SUSY versions of the Ma model 
and the AKS model, respectively.
We have also found out the benchmark point for each contributions,
where the neutrino oscillation data are correctly 
reproduced with satisfying the condition of strong 1stOPT and with 
satisfying the current experimental constraints.
Our model is a candidate of the fundamental theory whose 
low energy effective theory provides solutions to three 
serious problems in the SM; {\it i.e.}, neutrino mass, DM and baryogenesis
by physics at the TeV scale.
Our model can be tested at current and future collider experiments.

\begin{acknowledgments}
This work was supported in part by Grant-in-Aid for Scientific Research, Nos. 22244031 (S.K.), 
23104006 (S.K.), 23104011 (T.S.) and 24340046 (S.K. and T.S.). 
\end{acknowledgments}

\appendix
\section{Mass matrices and mixing matrices for extra fields}
Here we will list the mass terms of $Z_2$-odd particles which are obtained from 
the superpotential given by Eq.~(\ref{eq:superpotentialWeff}) and the soft SUSY breaking terms 
given by Eq.~(\ref{eq:softbrreakingterms}), and 
we will define the mixing matrices.

The mass terms for the $Z_2$ odd neutral scalars are given by
\begin{equation}
	\mathcal{L}=-
	\begin{pmatrix}
		\Phi_u^{\text{even}}&
		\zeta^{\text{even}}&
		\Phi_d^{\text{even}}&
		\eta^{\text{even}}&
		\Phi_u^{\text{odd}}&
		\zeta^{\text{odd}}&
		\Phi_d^{\text{odd}}&
		\eta^{\text{odd}}
	\end{pmatrix}
	M_0^2
	\begin{pmatrix}
		\Phi_u^{\text{even}}\\
		\zeta^{\text{even}}\\
		\Phi_d^{\text{even}}\\
		\eta^{\text{even}}\\
		\Phi_u^{\text{odd}}\\
		\zeta^{\text{odd}}\\
		\Phi_d^{\text{odd}}\\
		\eta^{\text{odd}}
	\end{pmatrix}\;,
\end{equation}
where the superscript "even" and "odd" denote the CP-even 
neutral scalar component and CP-odd neutral scalar component 
respectively.
The $8\times 8$ mass matrix $M_0^2$ can be written as
\begin{equation}
	M_0^2=
	\begin{pmatrix}
		M_{\varphi\varphi}^2&
		M_{\varphi\chi}^2\\
		(M_{\varphi\chi}^{2})^T&
		M_{\chi\chi}^2
	\end{pmatrix}\;,
\end{equation}
where the three $4\times 4$ matrices are defined as
\begin{align}
	M_{\varphi\varphi}^2=
	\mathrm{Re}M_{\phi^0}^2
	+\begin{pmatrix}
		0&0&0&0\\
		0&\mathrm{Re}(B_{\zeta}^2)&0&\mathrm{Re}(m_{\zeta\eta}^2)\\
		0&0&0&0\\
		0&\mathrm{Re}(m_{\zeta\eta}^2)&0&\mathrm{Re}(B_{\eta}^2)\\
	\end{pmatrix}
	\;,
\end{align}
\begin{align}
	M_{\chi\chi}^2=
	\mathrm{Re}M_{\phi^0}^2
	+\begin{pmatrix}
		0&0&0&0\\
		0&-\mathrm{Re}(B_{\zeta}^2)&0&\mathrm{Re}(m_{\zeta\eta}^2)\\
		0&0&0&0\\
		0&\mathrm{Re}(m_{\zeta\eta}^2)&0&-\mathrm{Re}(B_{\eta}^2)\\
	\end{pmatrix}\;,
\end{align}
\begin{align}
	M_{\varphi\chi}^2=
	-\mathrm{Im}M_{\phi^0}^2
	+\begin{pmatrix}
		0&0&0&0\\
		0&-\mathrm{Im}(B_{\zeta}^2)&0&-\mathrm{Im}(m_{\zeta\eta}^2)\\
		0&0&0&0\\
		0&\mathrm{Im}(m_{\zeta\eta}^2)&0&-\mathrm{Im}(B_{\eta}^2)\\
	\end{pmatrix}\;,
\end{align}
and 
\begin{align}
	M_{\varphi^0}^2=
	\begin{pmatrix}
		\bar{m}_{\Phi_u}^2+\hat{\lambda}^2\frac{v_d^2}{2}+D_{\Phi^0}&\hat{\lambda}^*\mu\frac{v_u}{\sqrt{2}}+A_{\zeta}^*\frac{v_d}{\sqrt{2}}&
		-B_{\Phi}^*\mu_{\Phi}^*&\hat{\lambda}^*\mu_{\Omega}\frac{v_d}{\sqrt{2}}-\hat{\lambda}\mu_{\Phi}^*\frac{v_u}{\sqrt{2}}\\
		\hat{\lambda}\mu^*\frac{v_u}{\sqrt{2}}+A_{\zeta}\frac{v_d}{\sqrt{2}}&\bar{m}_{\zeta}^2+\hat{\lambda}^2\frac{v_d^2}{2}&
		\hat{\lambda}\mu_{\Phi}^*\frac{v_d}{2}-\hat{\lambda}^*\mu_{\Omega}\frac{v_u}{\sqrt{2}}&B_{\Omega}\mu_{\Omega}\\
		-B_{\Phi}\mu_{\Phi}&\hat{\lambda}^*\mu_{\Phi}\frac{v_d}{\sqrt{2}}-\hat{\lambda}\mu_{\Omega}^*\frac{v_u}{\sqrt{2}}&
		\bar{m}_{\Phi_d}^2+\hat{\lambda}^2\frac{v_u^2}{2}-D_{\Phi^0}&-\hat{\lambda}\mu^*\frac{v_d}{\sqrt{2}}-A_{\eta}\frac{v_u}{\sqrt{2}}\\
		\hat{\lambda}\mu_{\Omega}^*\frac{v_d}{\sqrt{2}}-\hat{\lambda}^*\mu_{\Phi}\frac{v_u}{\sqrt{2}}&B_{\Omega}^*\mu_{\Omega}^*&
		-\hat{\lambda}^*\mu\frac{v_d}{\sqrt{2}}-A_{\eta}^*\frac{v_u}{\sqrt{2}}&\bar{m}_{\eta}^2+\hat{\lambda}^2\frac{v_u^2}{2}\\
	\end{pmatrix}\;.
\end{align}
The matrix $M_0^2$ is diagonalized by a real 
orthogonal matrix $O_0$ as 
\begin{equation}
	O_0^TM_0^2O_0^{}=
	\begin{pmatrix}
		m_{\Phi_1^0}^2&0&0&0&0&0&0&0\\
		0&m_{\Phi_2^0}^2&0&0&0&0&0&0\\
		0&0&m_{\Phi_3^0}^2&0&0&0&0&0\\
		0&0&0&m_{\Phi_4^0}^2&0&0&0&0\\
		0&0&0&0&m_{\Phi_5^0}^2&0&0&0\\
		0&0&0&0&0&m_{\Phi_6^0}^2&0&0\\
		0&0&0&0&0&0&m_{\Phi_7^0}^2&0\\
		0&0&0&0&0&0&0&m_{\Phi_8^0}^2
	\end{pmatrix}\;.
\end{equation}

The mass terms for $Z_2$-odd neutral fermions are written as
\begin{equation}
	\mathcal{L}=
	-\frac{1}{2}
	\begin{pmatrix}
		\tilde{\Phi}_u^0&
		\tilde{\zeta}^0&
		\tilde{\Phi}_d^0&
		\tilde{\eta}^0
	\end{pmatrix}
	\tilde{M}_0
	\begin{pmatrix}
		\tilde{\Phi}_u^0\\
		\tilde{\zeta}^0\\
		\tilde{\Phi}_d^0\\
		\tilde{\eta}^0
	\end{pmatrix}\;,
\end{equation}
where the mass matrix is given by 
\begin{equation}
	\tilde{M}_0
	=\begin{pmatrix}
		0&\hat{\lambda}\frac{v_d}{\sqrt{2}}&\mu_{\Phi}&0\\
		\hat{\lambda}\frac{v_d}{\sqrt{2}}&0&0&\mu_{\Omega}\\
		\mu_{\Phi}&0&0&-\hat{\lambda}\frac{v_u}{\sqrt{2}}\\
		0&-\hat{\lambda}\frac{v_u}{\sqrt{2}}&\mu_{\Omega}&0
	\end{pmatrix}\;.
\end{equation}
The mass matrix $\tilde{M}_0$ can be diagonalized by a unitary matrix $\tilde{U}_0$ as 
\begin{equation}
	\tilde{U}_0^T\tilde{M}_0\tilde{U}_0^{}=
	\begin{pmatrix}
		m_{\tilde{\Phi}_1^0}^{}&0&0&0\\
		0&m_{\tilde{\Phi}_2^0}^{}&0&0\\
		0&0&m_{\tilde{\Phi}_3^0}^{}&0\\
		0&0&0&m_{\tilde{\Phi}_4^0}^{}
	\end{pmatrix}\;,
\end{equation}
and one can obtain the real and positive mass eigenvalues $m_{\tilde{\Phi}_i}$.

The mass terms
for the $Z_2$-odd charged scalars are given by
\begin{equation}
	\mathcal{L}=-
	\begin{pmatrix}
		(\Phi_u^+)^*&
		(\Omega^+)^*&
		\Phi_d^-&
		\Omega^-
	\end{pmatrix}
	M_{\pm}^2
	\begin{pmatrix}
		\Phi_u^+\\
		\Omega^+\\
		(\Phi_d^-)^*\\
		(\Omega^-)^*\\
	\end{pmatrix}\;,
\end{equation}
with the mass matrix being
\begin{align}
	M_{\pm}^2=\begin{pmatrix}
		\bar{m}_{\Phi_u}^2+\hat{\lambda}^2\frac{v_u^2}{2}+D_{\Phi\pm}^{}&
		\hat{\lambda}\mu_{\Phi}^*\frac{v_d}{\sqrt{2}}-\hat{\lambda}^*\mu_{\Omega}\frac{v_u}{\sqrt{2}}&
		B^*\mu_{\Phi}^*&
		\hat{\lambda}^*\mu\frac{v_d}{\sqrt{2}}-A_{\Omega^-}^*\frac{v_u}{\sqrt{2}}\\
		\hat{\lambda}^*\mu_{\Phi}\frac{v_d}{\sqrt{2}}-\hat{\lambda}\mu_{\Omega}^*\frac{v_u}{\sqrt{2}}&
		\bar{m}_{\Omega^+}^2+\hat{\lambda}^2\frac{v_d^2}{2}+D_{\Omega\pm}^{}&
		-\hat{\lambda}^*\mu\frac{v_u}{\sqrt{2}}+A_{\Omega^+}^*\frac{v_d}{\sqrt{2}}&
		B^*\mu_{\Omega}^*\\
		B\mu_{\Phi}&
		-\hat{\lambda}\mu^*\frac{v_u}{\sqrt{2}}+A_{\Omega^+}^{}\frac{v_d}{\sqrt{2}}&
		\bar{m}_{\Phi_d}^2+\hat{\lambda}^2\frac{v_d^2}{2}-D_{\Phi\pm}^{}&
		\hat{\lambda}\mu_{\Omega}^*\frac{v_d}{\sqrt{2}}-\hat{\lambda}^*\mu_{\Phi}\frac{v_u}{\sqrt{2}}\\
		\hat{\lambda}\mu^*\frac{v_d}{\sqrt{2}}-A_{\Omega^-}\frac{v_u}{\sqrt{2}}&
		B\mu_{\Omega}&
		\hat{\lambda}^*\mu_{\Omega}\frac{v_d}{\sqrt{2}}-\hat{\lambda}\mu_{\Phi}^*\frac{v_u}{\sqrt{2}}&
		\bar{m}_{\Omega^-}^2+\hat{\lambda}^2\frac{v_u^2}{2}-D_{\Omega\pm}^{}
	\end{pmatrix}\;.
\end{align}
The mass matrix $M_{\pm}^2$ can be diagonalized by a unitary matrix $U_{+}^{}$ as 
\begin{equation}
	U_{+}^{\dagger}M_{\pm}^2{U}_{+}^{}=
	\begin{pmatrix}
		m_{{\Phi}_1^{\pm}}^{2}&0&0&0\\
		0&m_{{\Phi}_2^{\pm}}^{2}&0&0\\
		0&0&m_{{\Phi}_3^{\pm}}^{2}&0\\
		0&0&0&m_{{\Phi}_4^{\pm}}^{2}
	\end{pmatrix}\;.
\end{equation}

The mass terms of the 
$Z_2$-odd charged fermions
are written as 
\begin{equation}
	\mathcal{L}=
	-\begin{pmatrix}
		\tilde{\Phi}_u^+&\tilde{\Omega}^+
	\end{pmatrix}
	\tilde{M}_{\pm}
	\begin{pmatrix}
		\tilde{\Phi}_d^-\\
		\tilde{\Omega}^-
	\end{pmatrix}\;,
\end{equation}
where the mass matrix is given by
\begin{equation}
	\tilde{M}_{\pm}=\begin{pmatrix}
		-\mu_{\Phi}&\hat{\lambda}\frac{v_u}{\sqrt{2}}\\
		-\hat{\lambda}\frac{v_d}{\sqrt{2}}&-\mu_{\Omega}
	\end{pmatrix}\;.
\end{equation}
The mass matrix $\tilde{M}_{\pm}$ is diagonalized by two unitary matrices $U_L$ and $U_R$ as 
\begin{equation}
	U_R^{\dagger}\tilde{M}_{\pm}^{}U_L^{}=
	\begin{pmatrix}
		m_{\tilde{\Phi}^{\pm}_1}&0\\
		0&m_{\tilde{\Phi}^{\pm}_2}
	\end{pmatrix}\;,
\end{equation}
where $m_{\tilde{\Phi}_i^{\pm}}$ are the real and positive mass eigenvalues.

\end{document}